\documentclass[aps,prl,twocolumn,superscriptaddress,showpacs,floatfix]{revtex4}

\usepackage{graphicx,color,epsfig,rotate}

\begin{document}

\title {Quasi-one-dimensional magnons in an intermetallic marcasite}

\author{M. B. Stone}
\affiliation{Quantum Condensed Matter Division, Oak Ridge National
Laboratory, Oak Ridge, Tennessee 37831, USA}

\author{M. D. Lumsden}
\affiliation{Quantum Condensed Matter Division, Oak Ridge National
Laboratory, Oak Ridge, Tennessee 37831, USA}

\author{S. E. Nagler}
\affiliation{Quantum Condensed Matter Division, Oak Ridge National
Laboratory, Oak Ridge, Tennessee 37831, USA}

\author{D. J. Singh}
\affiliation{Materials Science and Technology Division, Oak Ridge National
Laboratory, Oak Ridge, Tennessee 37831, USA}

\author{J. He}
\affiliation{Department of Physics and Astronomy, Clemson University, Celmson, SC 29634, USA}

\author{B. C. Sales}
\affiliation{Materials Science and Technology Division, Oak Ridge National
Laboratory, Oak Ridge, Tennessee 37831, USA}

\author{D. Mandrus}
\affiliation{Materials Science and Technology Division, Oak Ridge National
Laboratory, Oak Ridge, Tennessee 37831, USA}
\affiliation{Department of Materials Science and Engineering, University of Tennessee, Knoxville, Tennessee 37996, USA}

\begin{abstract}
We present inelastic neutron scattering measurements and first principles calculations examining the intermetallic marcasite CrSb$_2$.  The observed spin wave dispersion implies that the magnetic interactions are strongly one-dimensional with antiferromagnetic chains parallel to the crystalline $c$-axis.   Such low-dimensional excitations are unexpected in a semiconducting intermetallic system.   Moreover this material may be further interesting in that the magnetic anisotropy may enhance thermoelectric properties along particular crystallographic directions.
\end{abstract}

\pacs{75.10.Jm,  
      75.40.Gb,  
      75.30.Et,  
      75.30.Ds, 
      78.70.Nx 
            }

\maketitle
Efficient and cost-effective power generation has revived interest in thermoelectric materials in an energy resource dominated economy.  CrSb$_2$ has gained recent attention as a narrow gap semiconductor with potential thermoelectric properties.  Investigations into substitutional compounds such as CrSb$_{2-x}$Te$_x$ \cite{Li_matsci_eng_2008,LiChina2010}, CrSb$_{2-x}$Sn$_x$\cite{LiChina2010} and Cr$_{1-x}$Ru$_x$Sb$_2$ \cite{takahashi2008} have found changes in the electronic properties, with an increase in the thermoelectric figure of merit for Te substitution.  CrSb$_2$ has also been considered as an electrode material for lithium based batteries showing improvement compared to using pure antimony.\cite{fernandez2001,park2010}  Here we report inelastic neutron scattering (INS) measurements and first principles calculations for this compound.  We find that CrSb$_2$ has significant exchange anisotropy exhibiting a quasi-one-dimensional (1d) magnon spectrum.  This spectrum has implications for the thermal conductivity as a function of crystallographic direction.  In addition, the determined exchange constants and anisotropy allow one to classify CrSb$_2$ as an $S=1$ antiferromagnetic (AFM) chain with interchain coupling and on-site anisotropy.

\begin{figure}[b]
\centering\includegraphics[scale=0.69]{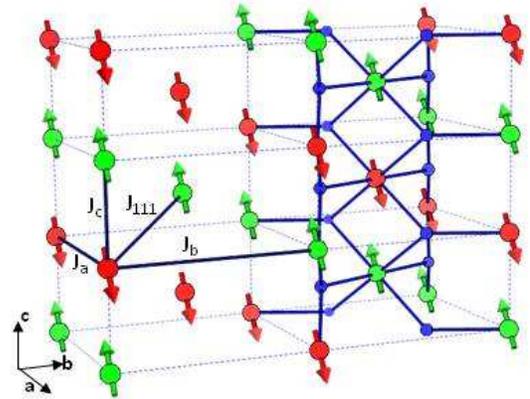}
\caption{\label{fig:structurefig}
(color online)Nuclear and magnetic structure of CrSb$_2$.\cite{holseth1968,holseth1970}  Red and green spheres are Cr sites, where the vector and color represent the spin orientation.  Small blue spheres are the Sb sites shown only for the three unit cells on the right of the figure for clarity.  Nearest neighbor exchange constants are labeled and shown as heavy black lines.}
\end{figure}

CrSb$_2$ is an intermetallic marcasite where the Cr atoms are located at the body center and corner positions of the orthorhombic unit cell with room temperature lattice constants $a=6.028$, $b=6.874$, and $c=3.272$~{\AA}.\cite{holseth1968}  The Cr atoms are at the center of a distorted octahedron of Sb sites as shown in Fig.~\ref{fig:structurefig}.  Early thermodynamic measurements found CrSb$_2$ to have AFM correlations\cite{foex1939,haraldsen1948}, but the ordered magnetic phase with $T_N=273(2)$~K was not observed until much later.\cite{holseth1970} The ordered magnetic structure is shown in Fig.~\ref{fig:structurefig} where the moments are roughly perpendicular to the (101) plane.  This magnetic structure results in magnetic Bragg peaks at the ($H$~$\frac{2K+1}{2}$~$\frac{2L+1}{2}$) positions where $H$, $K$, and $L$ are the Miller indices of the chemical unit cell.  The body centered Cr sites form the identical AFM structure as the corner sites, albeit displaced along the [111] direction.  The magnetic unit cell is doubled along the $b$ and $c$ axes with an ordered moment corresponding to spin $S=0.97(2)$ as determined from neutron diffraction measurements.\cite{holseth1970}  Early band structure considerations have shown that there are two localized d electrons per Cr$^{4+}$ in CrSb$_2$. \cite{goodenough}

Local spin density approximation (LSDA) calculations were performed using the general potential linearized augmented planewave (LAPW) method.\cite{singh}  Local orbital extensions were employed to include the high lying semi-core states and to relax any linearization errors.  The calculations included no shape approximations to either the potential or charge density.  Relativity was included for the valence states within a scalar relativistic approximation, while full relativity was included for the core states, within an atomic like approximation.  Well converged basis sets consisting of more than 850 LAPW functions plus local orbitals were used for the primitive unit cell (two formula units) of CrSb$_2$. Brillouin zone samplings were done with the special $\mathbf{k}$-points method, using 128 $\mathbf{k}$-points in the irreducible wedge of the primitive cell and equivalent samplings for the larger AFM cells.

\begin{figure}[b]
\centering\includegraphics[scale=0.89]{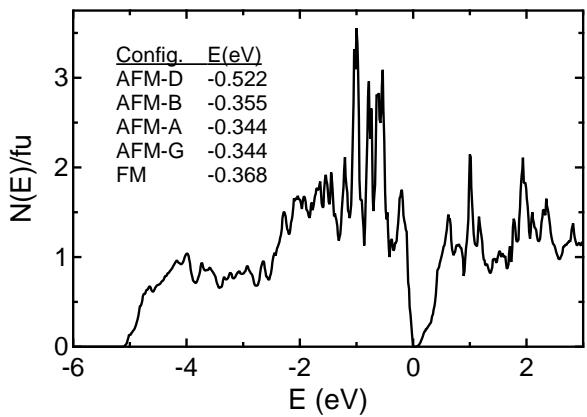}
\caption{\label{fig:dosfig}
LSDA calculated electronic density of states $N(E)$ per formula unit (fu) for the AFM order of the magnetic ground state (AFM-D).  The table inset lists the configurations and calculated LSDA energies per formula unit relative to the non-spin-polarized case (see text).}
\end{figure}

We consider ferromagnetic (FM) and several AFM ground states including the ordered magnetic structure found for CrSb$_2$, which we denote as AFM-D.  Other AFM structures include doubled magnetic cells along the $a$ (AFM-A) and $b$ (AFM-B) axes, and FM correlations within each of the sublattices but AFM relative to one another (AFM-G).  The lowest calculated LSDA energy is found for AFM-D, -0.522 eV per formula unit relative to the non-spin polarized case, with other values listed in Fig.~\ref{fig:dosfig}.  All of the configurations were found to be conducting except for AFM-D.  The calculated density of states for AFM-D is shown in Fig.~\ref{fig:dosfig}.  The calculated bandgap of 0.07 eV agrees with low temperature resistivity measurements.~\cite{adachi1969}  The large energy difference between the AFM-D and FM structures compared to the relatively low $T_N$ in a lattice with no geometrical frustration suggests interesting magnetic correlations.  The energy of the AFM-G state is 0.025 eV per Cr higher than the FM state, indicating that the inter-sublattice coupling, $J_{111}$ in Fig.~\ref{fig:structurefig}, is very weak and likely FM.  The energy of the other AFM states implies weak FM exchange along $a$, almost no interaction along $b$, and strong AFM exchange along $c$ in agreement with the ordered magnetic structure.  The large AFM exchange along the $c$ axis would make the magnetic correlations in CrSb$_2$ quasi-1d, thus explaining the suppressed value of $T_N$.\cite{scalapino}  This also explains the difficulty in observing the phase transition in prior thermodynamic measurements.  The 1d magnetism would result in much of the magnetic entropy being exhausted with a very broad hump in susceptibility as the local order develops, with only smaller features at $T_N$.

Single crystal INS provides an efficient probe of the LSDA predictions of the exchange coupling for CrSb$_2$.  Single crystals were grown from an Sb flux with a starting composition of CrSb$_{16}$. The material was heated in vacuum in a 10~cc alumina crucible to 1273~K, held for 24~h, and then cooled to 913~K at 1~K/hr. The excess Sb liquid was centrifuged away, leaving crystals with typical dimensions of 5x5x5~mm$^3$. Single crystal INS measurements of CrSb$_2$ were performed at the HB3 triple-axis spectrometer (TAS) and the ARCS chopper spectrometer at Oak Ridge National Laboratory's High Flux Isotope Reactor and Spallation Neutron Source respectively.  Measurements at HB3 were performed for samples in the (0 K L)[$m\approx 4.05$~g] and (H L L)[$m\approx 3.5$~g] scattering planes.  The instrument was configured with a vertically focused Be (002) monocrhomator and a PG (002) analyzer with fixed final energy $E_f=30.5$. Pyrolytic graphite filters were in place between the sample and analyzer.  Constant wave vector, $\mathbf{Q}$, and energy transfer, $\hbar\omega$, scans were performed primarily along the (H~$\kappa/2$~$\kappa/2$), (0~K~$\kappa/2$) and (0~$\kappa/2$~L) directions, where $\kappa = 1$~or~$3$ between 5 and 70 meV.  Data were acquired for a fixed monitor count unit (mcu) to account for variations in neutron flux as a function of incident energy, $E_i$.  Measurements at ARCS were performed with the (0 K L) plane in the equatorial plane of the instrument with $E_i =150$~meV neutrons.  Data were measured by rotating about the vertical (1 0 0) axis in increments of 0.75 degrees over a total of 62 degrees with the incident wave vector initially pointed 9 degrees from the (0 1 0) direction.  Data were normalized to the accumulated charge on the source and histogramed into $\mathbf{Q}$ and $\hbar\omega$ space.

\begin{figure*}[t!]
\centering\includegraphics[scale=0.89]{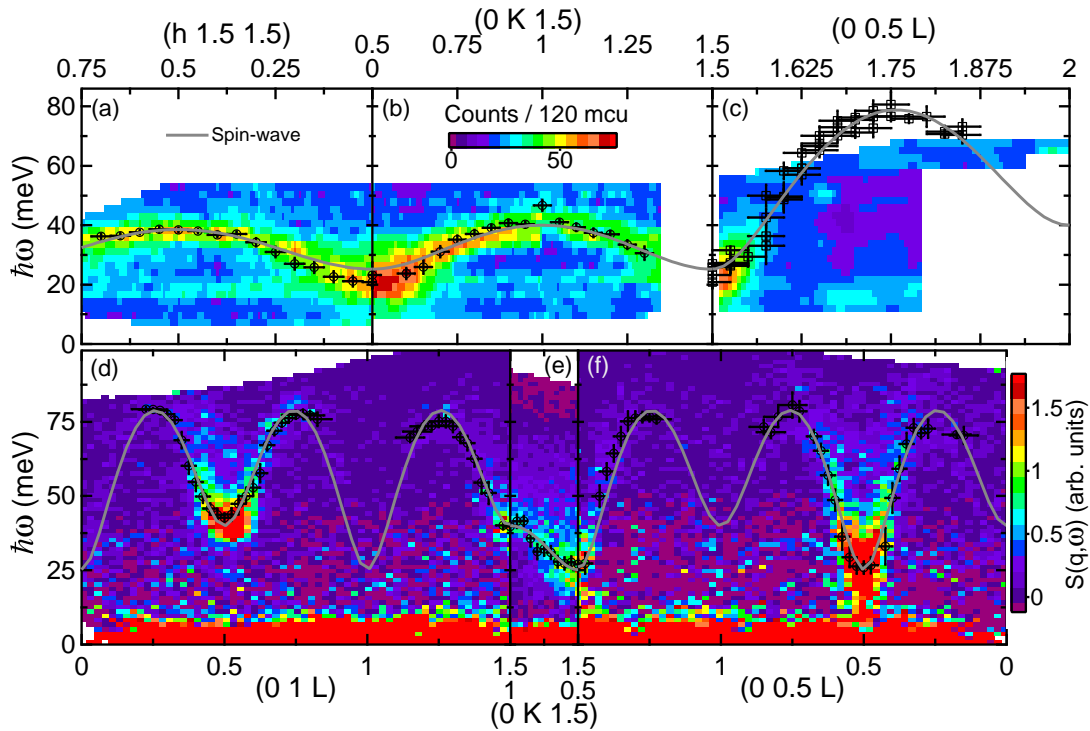}
\caption{\label{fig:dispersion}(Color online) $T = 10$~K INS spectra of CrSb$_2$ along principal axes of the reciprocal lattice.  (a)-(c)[(d)-(f)] Data were measured using HB3[ARCS]. (d)-(f) Have been background subtracted as described in the text.
(a)[(b)-(f)] were measured with the sample in the (H K K)[(0 K L)] scattering plane.  Open circles correspond to peak positions determined from extracted constant $\mathbf{Q}$ scans as described in the text.  Open squares in (c) correspond to peak positions from ARCS and HB3 constant $\mathbf{Q}$ scans along the ($H$~$\frac{(2K+1)}{2}$~L) directions where $H$ and $K$ are integers.  Vertical[Horizontal] error bars represent the fitted peaks full width at half maximum [range of $\mathbf{Q}$ integration].  The grey solid line is the fitted spin-wave dispersion described in the text.}
\end{figure*}

Figure~\ref{fig:dispersion} shows the measured scattering intensity as a function of $\hbar\omega$ and $\mathbf{Q}$ for CrSb$_2$ at $T=10$~K.  Given the inversion symmetry of the magnetic structure, the data from ARCS, Fig.~\ref{fig:dispersion}(d)-(f), have been averaged for data acquired at $\pm\mathbf{Q}$ and for higher $\mathbf{Q}$ Brillouin zones scaled by the Cr magnetic form factor.  Data have been background subtracted using data scaled from larger Brillouin zones to suppress phonon scattering.  There is a clear single excitation that disperses along all three principal axes.  The overall minimum in the dispersion appears at $\approx25$~meV at $(H,\frac{2K+1}{2},\frac{2L+1}{2})$~for integer $H$,$K$ and $L$.  The kinematic limits of the TAS measurements did not allow us to follow this mode above 60 meV for small values of $|\mathbf{Q}|$.  The ARCS measurements are able to observe this mode up to its maximum energy of $\approx 78$~meV.  The bandwidth along the $H$ and $K$ direction is 13.3 and 14.8 meV respectively, and the bandwidth of the excitation along $L$ is 53.6 meV.

\begin{figure}[t]
\centering\includegraphics[scale=0.89]{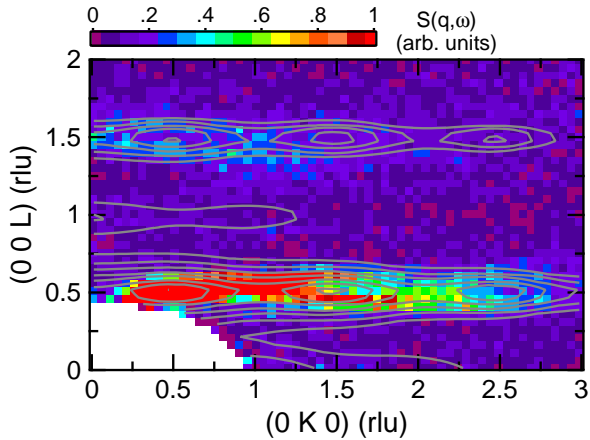}
\caption{\label{fig:iofqfig}
(color online)Intensity in the (0KL) plane integrated for $25\leq \hbar\omega \leq 100$~meV and $|H|<0.1$~rlu (reciprocal lattice units).  Data are from the ARCS measurements at $T=10$~K.  Solid lines are the contours of the calculated spin-wave scattering intensity as described in the text.  Contour lines are plotted in logarithmic steps.  White area in bottom left was not sampled during the measurement.}
\end{figure}

Figure~\ref{fig:iofqfig} shows the (0KL) scattering intensity from the ARCS measurement integrated for $25\leq \hbar\omega \leq 100$~meV.  The scattering intensity falls off rapidly with $|\mathbf{Q}|$ indicating magnetic fluctuations.  There are ridges of scattering consistent with the greatest dispersion being along the $c$-axis as shown in Fig.~\ref{fig:dispersion}.
Prior investigations of the high temperature magnetic susceptibility employed a two-state magnetic system with an energy gap of 70.7 meV to describe the results.\cite{adachi1969}  Although this value is within the energy scale of the measured magnon, it is quite far from the peak in the magnetic density of states.  Our INS measurements provide a picture of the microscopic interactions in this material, and clearly shows that the gap in the magnetic spectrum is approximately 25 meV.

We project the data from ARCS and HB3 into a collection of constant $\mathbf{Q}$ scans by integrating over $\pm0.05$ rlu in each of the principal axes directions.  These individual scans are then fit to a single inelastic Gaussian peak for $20<\hbar\omega <80$~meV, allowing us to extract the dispersion of the magnetic excitation.  A portion of these fitted peak positions is shown in Fig.~\ref{fig:dispersion}.  The points in Fig.~\ref{fig:dispersion}(c) are from a series of $\mathbf{Q}$'s along the ($\eta$~$\frac{2\kappa+1}{2}$~L) direction where $\eta$ and $\kappa$ are integer values measured with both HB3 and ARCS.  This panel illustrates the extent of the dispersion along the $L$ direction.  Although the electronic band-gap is small, we consider the cooperative magnetism in CrSb$_2$ to be local moment in nature.  This allows one to use linear spin wave theory (LSWT) to calculate a  dispersion relation, $\hbar\omega(\mathbf{Q})_{SW}$ for comparison to the measured data.  The tilts and distortion of the Cr-Sb octahedron, the direction of the ordered moments, and the gap in the AFM spectrum implies the need for an anisotropy parameter, $D$. The Heisenberg Hamiltonian then becomes
\begin{equation}
\label{eq:hamiltonian}
\displaystyle\mathcal{H} = \sum\limits_{\alpha}\sum\limits_{\langle ij\rangle_{\alpha}}J_\alpha\mathbf{S}_{i}\cdot\mathbf{S}_{j} + \sum\limits_{i} D(S_z^2)_i,
\end{equation}
where $\alpha$ is summed over $a$, $b$, and $c$, and we define the $z$ axis to be along the ordered moment.\cite{lovesey_vol2,otherspinwave}  This results in the spin-wave dispersion
\begin{eqnarray}
\label{eq:swdispeqn}
\nonumber A_{\mathbf{Q}} &=& 2S[J_a[\cos(2\pi H)-1]+J_b + J_c + D] \\
\nonumber B_{\mathbf{Q}} &=& 2S[J_b\cos(2\pi K) + J_c\cos(2\pi L) ]
\\
\hbar\omega(\mathbf{Q})_{SW} &=& \sqrt{A_{\mathbf{Q}}^2 - B_{\mathbf{Q}}^2},
\end{eqnarray}
where $S$ is the magnitude of the spin quantum and the exchange constants are illustrated in Fig.~\ref{fig:structurefig}.    Fitting to $\hbar\omega(\mathbf{Q})_{SW}$ with $S=1$ results in the solid grey line in Fig.~\ref{fig:dispersion} and the parameters $J_c = 35.7(2)$, $J_a=-1.31(6)$, $J_b = 1.69(7)$ and $D=2.07(9)$~meV.  Including a $J_{111}$ exchange constant has negligible improvement on the $\chi^2$ of the fit.  LSWT accounts well for the measured low-temperature dispersion in CrSb$_2$.  The resulting values also agree well with the FM and AFM spin-spin interactions based upon the low temperature ordered magnetic structure, and the predictions based upon LSDA calculations discussed earlier.  The $J_c$ exchange is an order of magnitude larger than along the other axes in support of the predicted quasi-1d magnon behavior.

We also examine the integrated scattering intensity in the $(0KL)$ plane as shown in Fig.~\ref{fig:iofqfig}.  We calculate the energy resolution convolved one-magnon cross-section \cite{lovesey_vol2} based upon the determined exchange constants and the resolution function of the ARCS spectrometer.  We include the Cr magnetic form factor, and plot the cross section as a contour in Fig.~\ref{fig:iofqfig}.  The calculated scattering intensity agrees very well with the measured spectrum.

For $T > T_N$, the magnetic excitations in CrSb$_2$ could be considered $S=1$ quasi-1d magnons propagating out of a spin-liquid ground state suggesting comparison to a Haldane chain dispersion.  Thermal population of phonons makes an accurate comparison using INS difficult at high temperatures, but the extracted exchange constants based upon LSWT for $T<T_N$ place CrSb$_2$ well within an ordered phase of the phase diagram.\cite{haldanephasediag1,haldanephasediag2} Measurements of the narrow-gap semiconducting marcasite FeSb$_2$ have found this system to be nearly magnetic, and that substitution with Cr as Fe$_{1-x}$Cr$_x$Sb$_2$ with $x \geq 0.45$ pushes the system into the AFM phase with $T_N$ increasing with $x$.\cite{petrovic1,petrovic2}  Although INS of FeSb$_2$ observed no clear magnetic excitations,\cite{zaliznyak} investigations into suppressing the $T_N$ of CrSb$_2$ via chemical substitution are fundamentally interesting to try to move through the Haldane phase diagram and ultimately access the spin-liquid Haldane phase.  Another route to adjusting the exchange constants along with the electrical conductivity would be to change the atomic percent of Sb.  The high temperature trend of the magnetic susceptibility is very similar for the range of compositions between CrSb$_{1.75}$ to CrSb$_{5}$.\cite{haraldsen1948}

The quasi-1d magnetic excitation spectrum of CrSb$_2$ will affect its thermoelectric properties.  Thermal transport measurements have been performed for other low-d magnets.  In the case of the $S=\frac{1}{2}$ spin-ladder Sr$_{14-x}$Ca$_x$Cu$_{24}$O$_41$\cite{telephone1} and the $S=1$ Haldane chain NENP\cite{NENP1}, it was found that thermal transport was enhanced along the 1d axis by a factor of up to $\approx 15$.  The large magnetic gap and spin-wave velocity found for CrSb$_2$ would serve to enhance the thermal conductivity along the 1d axis for temperatures of the order of $J_c$.  This implies that the room temperature thermoelectric figure of merit may be improved by orienting the 1d axis orthogonal to the flow of charge.  While the lattice thermal conductivity is probably dominant, the magnetic contribution is potentially important and it along with its anisotropy should be considered when assessing the thermoelectric properties of CrSb$_2$.

MBS acknowledges valuable discussions with I. Zaliznyak.  Research at Oak Ridge National Laboratory's High Flux Isotope Reactor and Spallation Neutron Source was sponsored by the Scientific User Facilities Division, Office of Basic Energy Sciences, U. S. Department of Energy.  DJS, BCS and DM supported by the Department of Energy office of  Materials Science and Engineering.  MBS, MDL, and SEN supported by the Department of Energy office of Scientific User Facilities.  JH would like to acknowledge the financial support from DOE grant No. DE-FG02-04ER-46139.


\begin{thebibliography}{99}
\bibitem{Li_matsci_eng_2008}
H. J. Li \textit{et al.}, Mat. Sci. Eng. B \textbf{149}, 53 (2008).

\bibitem{LiChina2010}
H. Li \textit{et al.}, Chi. J. Mat. Res. \textbf{24}, 429 (2010).

\bibitem{takahashi2008}
Y. Takahashi \textit{et al.}, J. Alloys Compds. \textbf{459}, 78 (2008).


\bibitem{fernandez2001}
F. J. Fern\'{a}ndez-Madrigal \textit{et al.}, J. Electro. Chem. \textbf{501}, 205 (2001).

\bibitem{park2010}
C-M. Park and H. J. Sohn, Electrochimica Acta, \textbf{55}, 4987 (2010).



\bibitem{holseth1968}
H. Holseth and A. Kjekshus, Acta Chem. Scand., \textbf{22}, 3283 (1968).

\bibitem{holseth1970}
H. Holseth, \textit{et al.}, Acta Chem. Scand., \textbf{24},
3309 (1970).

\bibitem{foex1939}
G. Fo\"{e}x and M. Graff, Compt. rend. \textbf{209}, 160 (1939).

\bibitem{haraldsen1948}
H. Haraldsen, \textit{et al.}, Arch. Math. Naturvidenskab, \textbf{50}, No. 4., 95-135 (1948).


\bibitem{goodenough}
J. B. Goodenough, J. Solid State Chem., \textbf{5}, 144 (1972).


\bibitem{singh}
D. J. Singh, \textit{Planewaves, Pseudopotentials and the LAPW method}, Kluwer Academic, Boston (1994).


\bibitem{adachi1969} K. Adachi, \textit{et al.}, J. Phys. Soc. Jpn.
\textbf{26}, 906 (1969).

\bibitem{scalapino}
D. J. Scalapino, \textit{et al.}, Phys. Rev. B \textbf{11}, 2042 (1975).



\bibitem{lovesey_vol2}
S. W. Lovesey, \textit{Theory of Neutron Scattering from Condensed Matter:  Volume 2} (Clarendon Press, Oxford, UK 1987).

\bibitem{otherspinwave}
P. A. Lindg{\aa}rd, \textit{et. al}, J. Phys. Chem. Solids \textbf{28}, 1357 (1967).

\bibitem{haldanephasediag1}
T. Sakai and M. Takahashi, Phys. Rev. B \textbf{42} 4537 (1990).

\bibitem{haldanephasediag2}
A. Zheludev, \textit{et al.}, Phys. Rev. B \textbf{62}, 8921 (2000).

\bibitem{petrovic1}
C. Petrovic, \textit{et al.}, Phys. Rev. B \textbf{67}, 155205 (2003).

\bibitem{petrovic2}
R. Hu, \textit{et al.}, Phys. Rev. B \textbf{76}, 115105 (2007).

\bibitem{zaliznyak}
I .A. Zaliznyak, \textit{et al.}, Phys. Rev. B \textbf{83}, 184414 (2011).

\bibitem{telephone1}
A. V. Sologubenko, \textit{et al.}, Phys. Rev. Lett. \textbf{84}, 2714 (2000).

\bibitem{NENP1}
A. V. Sologubenko, \textit{et al.}, Phys. Rev. Lett. \textbf{100}, 137202 (2008).
\end{thebibliography}
\end{document}